\documentstyle[12pt,aasms4]{article}
\begin{document}
\title{SOUND SPEED IN A RANDOM FLOW AND TURBULENT SHIFTS OF THE SOLAR EIGENFREQUENCIES}
\author{Andrei V. Gruzinov}
\affil{Institute for Advanced Study, School of Natural Sciences,
Princeton, NJ 08540}

\begin{abstract}
Perturbation theory is used to calculate the frequency shift of acoustic modes of an homogeneous turbulent fluid and the frequency shifts of solar modes due to turbulent convection. For sound waves in a random flow, the  fractional frequency shift is $+(11/30)M^2$ in the long-wavelength limit, and the shift at short wavelengths is $-(2/3)M^2$, where $M$ is the average Mach number of the flow. In our model of the solar convection, the low-degree f-mode shift is positive and unobservably small, whereas the fractional frequency shift at high degrees ($l\gg 500$) is $-M^2$, where $M$ is the Mach number of convection near the solar surface.
\end{abstract}
\eject

\section{INTRODUCTION}

Frequencies of low-order ($n\sim 1$) high-degree ($l\sim 1000$) solar modes are measured with an accuracy of about 0.2\% (Libbrecht, Woodard \& Kaufmann 1990, Bachmann, Duval, Harvey, \& Hill 1995). It is found that the observed frequencies of the fundamental mode are smaller than the theoretically expected value of $\sqrt{gk}$ by $\sim 1\% $. One may expect that turbulent convective flows with characteristic velocities $v\sim 1km/s$, in the region with a characteristic sound speed $c\sim 10km/s$, will cause a fractional frequency shift of the order $v^2/c^2$, $\sim 1\% $ shift. Magnetic fields seem to be less important than turbulent motions, because they reach equipartition strengths only in sunspots. Also, frequency shifts due to magnetic fields should be positive. We will ignore magnetic fields in what follows.

The negative sign of the shifts was explained by the following simple argument (Murawski \& Roberts 1993; Rosenthal 1996). Consider a sound wave propagation through a layer of thickness $2L$. Let the flow velocity along the wavevector be $+V$ in the first sub-layer of thickness $L$, and $-V$ in the second sub-layer. Then the total propagation time $t=L/(c+V)+L/(c-V)$, and the effective sound speed is $c_{\ast }=2L/t=(c^2-2V^2)^{1/2}$. Therefore the fractional frequency shift, $s=\delta \omega /\omega$, is 
\begin{equation}
s=-{1\over 3}M^2, 
\end{equation}
where $M=<V^2>/c^2$, and $1/3$ comes from the angle average. 

The above kinematic explanation is incomplete. The sound speed in a random homogeneous isotropic flow depends on the wavelength. We show in \S 2 that the effective sound speed is increased in the long wavelength limit and decreased in the short wavelength limit. We discuss the solar frequency shifts in \S 3. Our perturbation theory (PT) calculations are in principle straightforward, but algebraically tedious; appendices explain how we performed the calculations and tested the results.

\section {SOUND SPEED IN A HOMOGENEOUS ISOTROPIC RANDOM FLOW}

We will assume that the turbulence Mach number is so small that the turbulent flow is approximately stationary (even at time scales of the long wavelength acoustic modes). Only the turbulent velocity field, $V$, can give frequency shifts that are quadratic in $M$. Turbulent pressure, density and entropy fluctuations are quadratic in $M$, and give rise to $O(M^4)$ frequency shifts that can be neglected. To within this accuracy, $\nabla \cdot {\bf V}=0$ in the absence of stratification.

In the leading order, one can find the shift caused by a single Fourier harmonic of the flow, and then average the result over all harmonics. Since velocity gradient of a Fourier harmonic of the incompressible flow is perpendicular to the velocity, it is sufficient to solve the problem for a shear flow 
\begin{equation}
{\bf V}=(0,V(x),0).
\end{equation}
The sound wave frequency in an isotropic random flow is then obtained by taking an appropriate average. We will use the fact that $V$ is a pure Fourier harmonic later in our calculation, but at this stage $V$ may be thought of as an arbitrary shear flow with a zero mean.

\subsection{The Basic Equation}

We first obtain a convenient representation for the linear sound wave equation in the presence of a shear flow. Let the velocity perturbations ($u$, $v$, $w$) and dimensionless density perturbation ($\rho$) associated with the sound wave depend on $x$ in an unspecified way, and be proportional to $\exp (-i\omega t+ik_yy+ik_zz)$. Then, denoting the $x$-derivative by a prime, and using the standard notation $\Omega (x)=\omega -k_yV(x)$, we have
\begin{equation}
-i\Omega u=-c^2\rho ',
\end{equation}
\begin{equation}
-i\Omega v+V'u=-ic^2k_y\rho,
\end{equation}
\begin{equation}
-i\Omega w=-ic^2k_z\rho,
\end{equation}
\begin{equation}
-i\Omega \rho +u'+ik_yv+ik_zw=0.
\end{equation}

Equations (2)-(5) give the velocity perturbations in terms of the density perturbation. Using these expressions in Eq. (6), we get
\begin{equation}
({\Omega ^2\over c^2} -k_{\perp }^2)\rho +\rho '' -{2\Omega '\over \Omega }\rho '=0,
\end{equation}
where $k_{\perp }^2=k_y^2+k_z^2$. Up to the second order in $V$, this equation can be written as
\begin{equation}
-\rho ''+Q_1\rho +Q_2\rho=({\omega ^2\over c^2} -k_{\perp }^2)\rho ,
\end{equation}
where $Q_1$ is the linear perturbation operator
\begin{equation}
Q_1={2k_y\over ck}(k^2V-V'\partial ),
\end{equation}
and $Q_2$ is the quadratic perturbation operator
\begin{equation}
Q_2=-{k_y^2\over c^2k^2}(k^2V^2+2VV'\partial ).
\end{equation}
In the above expressions we have replaced $\omega$ by the unperturbed value $ck$, knowing that the frequency shift due to the linear perturbation $Q_1$ vanishes in the first order PT. 

\subsection{Perturbation Theory}

Quadratic perturbation $Q_2$ gives a finite shift in the first order PT. Denoting the unperturbed mode by $|k_x>=\exp (ik_xx)$, we have
\begin{equation}
\delta _2\omega ^2=c^2<k_x|Q_2|k_x>=-<V^2>k_y^2.
\end{equation} 
Averaging over angles gives a fractional frequency shift 
\begin{equation}
s_2=-{1\over 6}M^2,
\end{equation}
where $M^2=<V^2>/c^2$.

The linear perturbation $Q_1$ gives a non-zero frequency shift in the second order PT,
\begin{equation}
\delta _1\omega ^2=c^2\sum_{k'} {<k_x|Q_1|k'><k'|Q_1|k_x>\over k_x^2-k'^2}.
\end{equation} 
For a sinusoidal shear flow with a wavenumber $q$, the sum is over $k'=k_x\pm q$, and the above expression gives
\begin{equation}
\delta _1\omega ^2=4<V^2>({k_y^2\over k^2}) {k_{\perp}^2q^2-k^4\over q^2-4k_x^2}.
\end{equation} 
Averaging this equation over the angle between $k$ and $q$ yields
\begin{equation}
s_1=M^2\int_{0}^1 da (1-a^2){(1-a^2)q^2-k^2\over q^2-4a^2k^2},
\end{equation}
where $a$ is the cosine of the angle.
For $k>q/2$, there is a pole due to the resonant scattering. The frequency shift is given by the principal value of the integral. The imaginary part gives the damping, which in the solar context corresponds to the line-width (Goldreich \& Murray 1994). We do not consider line-widths in this work.

\subsection{Results}

The integral is trivial in the long-wavelength limit, $k\ll q$,
\begin{equation}
s_1=(8/15)M^2.
\end{equation} 
Summing the two contributions, Eq.(12) and Eq.(16), we get a total fractional frequency shift 
\begin{equation}
s={11\over30}M^2.
\end{equation}
In the short-wavelength limit, $k\gg q$, Eq.(15) gives
\begin{equation}
s_1=-{1\over2}M^2,
\end{equation} 
and the total short-wavelength shift is given by
\begin{equation}
s=-{2\over3}M^2.
\end{equation}

The fractional shifts in the asymptotic regions are wavevector independent. This is equivalent to the sound speed renormalization. The long-wavelength sound speed is higher than $c$, $c_{\ast }^2=c^2+{11\over15}<V^2>$, and the short-wavelength sound speed is lower than $c$, $c_{\ast }^2=c^2-{4\over3}<V^2>$. These results were confirmed analytically (Appendix A) and by numerical calculations (Appendix B).

\section {THE SOLAR FREQUENCY SHIFTS}

The above results are not directly applicable to the problem of the solar $p$ and $f$-mode frequency shifts due to interactions with  turbulent convection. The Sun is stratified, and the solar convection is anisotropic and inhomogeneous. However the method of \S 2, second-order PT analysis of the density equation, is still applicable.

In \S 3.1 we derive the basic density equation analogous to Eq.(8). In \S 3.2 we specify the model of the solar stratification in the upper convection zone. We give a general formula for the frequency shifts in \S 3.3, and calculate the f-mode shifts for a particular model of the solar convective turbulence in \S 3.4.

\subsection{The Basic Equation}

The aim is to calculate the eigenfrequencies of linear sound waves in a medium with an unperturbed density $\rho _0$, sound speed $c$, gravitational acceleration ${\bf g}$ in the presence of a flow ${\bf V}$. We use a planar model for which $\rho _0$ and $c$ depend on the depth $x$, whereas ${\bf g}$ is independent of the depth. The convective flow ${\bf V}$ satisfies the mass continuity equation $\nabla \cdot ({\bf V}\rho _0)=0$.

Velocity perturbations $v_{\alpha}$ and a density perturbation $\rho$ (unlike \S 2.1 $\rho$ is not normalized to $\rho _0$) associated with the acoustic mode satisfy the linearized equations
\begin{equation}
-i\omega v_{\alpha}+V_{\beta}\partial _{\beta}v_{\alpha}+v_{\beta}\partial _{\beta}V_{\alpha}=-{1\over \rho _0}\partial _{\alpha}c^2\rho+{1\over \rho _0}g_{\alpha}\rho,
\end{equation}
\begin{equation}
-i\omega \rho +\partial _{\beta}V_{\beta}\rho+\partial _{\alpha}\rho _0v_{\alpha}=0.
\end{equation}
Here and in what follows we use the convention that an operator acts on everything to the right of it, unless there is a bracket. We change variables, $\rho _0v_{\alpha}\rightarrow v_{\alpha}$, and define three operators
\begin{equation}
F_{\alpha}=\partial _{\alpha}c^2-g_{\alpha},
\end{equation}
\begin{equation}
G=\partial _{\alpha}V_{\alpha},
\end{equation}
\begin{equation}
H_{\alpha \beta}=\delta _{\alpha \beta}G+(\partial _{\beta}V_{\alpha}).
\end{equation}
Then the system (20), (21), can be written as
\begin{equation}
-i\omega v_{\alpha}+H_{\alpha \beta}v_{\beta}+F_{\alpha}\rho =0,
\end{equation}
\begin{equation}
-i\omega \rho+G\rho +\partial _{\alpha}v_{\alpha}=0.
\end{equation}
Now solve Eq.(25) for $v$ in terms of $\rho$ expanding up to the second order in $V$,
\begin{equation}
v_{\alpha}={1\over i\omega }F_{\alpha}\rho +{1\over (i\omega )^2}H_{\alpha \beta}F_{\beta}\rho+{1\over (i\omega )^3}H_{\alpha \beta}H_{\beta \gamma}F_{\gamma}\rho .
\end{equation}
Substituting this expansion into Eq.(26), we get the basic perturbed density equation (analogous to Eq.(8))
\begin{equation}
Q_0\rho +Q_1\rho +Q_2\rho=\omega ^2\rho .
\end{equation}
Here the unperturbed operator is
\begin{equation}
Q_0=-\partial _{\alpha}F_{\alpha},
\end{equation}
the linear perturbation operator is 
\begin{equation}
Q_1={i\over \omega}\partial _{\alpha}H_{\alpha \beta}F_{\beta}-i\omega G,
\end{equation}
and the quadratic perturbation operator is
\begin{equation}
Q_2={1\over \omega ^2}\partial _{\alpha}H_{\alpha \beta}H_{\beta \gamma}F_{\gamma}.
\end{equation}

Before we can use the PT technique on Eq. (28), we have to specify the stratification, find the unperturbed eigenmodes, and ``Hermitize'' $Q_0$.

\subsection{The Stratification in the Solar Convection Zone}

We use a polytropic model of the solar convection zone (see Goldreich \& Kumar 1990). In this model the unperturbed (bare) sound speed is $c^2=gx/m$, where $x$ is the depth and $m\approx 4$ is the polytropic index in the hydrogen ionization zone. Then the unperturbed density equation, Eq. (28) with $Q_1=Q_2=0$, reads
\begin{equation}
-{x\over m}(\rho ''-k^2\rho)+{m-2\over m}\rho '={\omega ^2\over g}\rho ,
\end{equation}
for a horizontal wavenumber $k$. An eigenmode with $n$ nodes in the vertical direction is of the form $\rho ^{(n)}=\exp (-kx)x^{m-1} L_n(kx)$, where $L_n$ is the Laguerre polynomial of degree $n$. The dispersion law is $\omega _n^2=(1+2n/m)gk$. The modes with $n>0$ are the p-modes, and the $n=0$ mode is the fundamental mode (f-mode). 

The fundamental mode frequency does not depend on the value of the polytropic index $m$, $\omega ^2=gk$. This a familiar dispersion law of the deep-water gravity waves. A remarkable fact is that the f-mode frequency is $\sqrt{gk}$ for an arbitrary stratification. The solar stratification in the upper convection zone is not well known (Christensen-Dalsgaard 1996), making interpretation of the p-mode shifts difficult. But the f-mode shifts must be mostly turbulent; we will concentrate on the f-mode shifts in \S 3.4.

For a PT to be applicable, the unperturbed operator, $Q_0$, must be Hermitian. This is achieved by introducing the scalar product
\begin{equation}
<\rho _1|\rho _2>=\int d^3rx^{1-m}\bar {\rho _1}\rho _2.
\end{equation}

Knowing the unperturbed solutions and the scalar product we can now calculate the frequency shifts.

\subsection {Frequency Shifts}

We can use the unperturbed frequency in the perturbation operators $Q_1$ and $Q_2$. The quadratic perturbation $Q_2$ gives a non-zero frequency shift in the first order PT. For a horizontal wavenumber $k$ and a mode order $n$, we have
\begin{equation}
\delta _2\omega _n^2(k)=<k,n|Q_2|k,n>.
\end{equation}
A non-zero frequency shift from $Q_1$ appears in the second order PT,
\begin{equation}
\delta _1\omega _n^2(k)=\sum_{k',n'}{<k,n|Q_1|k',n'><k',n'|Q_1|k,n>\over \omega _n^2(k)-\omega _{n'}^2(k')}.
\end{equation}
The total frequency shift is a sum of Eqs. (34) and (35). It remains to specify the flow $V$ and do the integrals.

\subsection {Turbulent Shifts of the Solar Fundamental Mode} 

We use a two dimensional model with a $y$-axis in the horizontal direction. For a single Fourier harmonic $\exp (iqy)$ of the turbulent flow $V$, the incompressibility of the mass flow, $\partial _{\alpha}(V_{\alpha}\rho _0)=0$, is satisfied by the velocity field of the form
\begin{equation}
V_x=-iq\Psi ,
\end{equation} 
\begin{equation}
V_y=\Psi '+{m\over x}\Psi .
\end{equation}
Since the sound speed scales as $x^{1/2}$, we choose the stream function that scales as $x^{3/2}$ at small depths $x$, so as to make the Mach number of the turbulence finite. We suppose that for a given horizontal scale of the convection, only one row of convective cells exists in the vicinity of the surface. This means that $\Psi$ has a maximum at the depth of the order of $1/q$, and then goes to zero at larger depths. We choose
\begin{equation}
\Psi =Ax^{3/2}e^{-\mu qx},
\end{equation}
where the $y$-dependence is omitted, $\mu$ measures the aspect ratio of the convective cells. We will sum the frequency shifts due to different Fourier components of the turbulence at the end of the calculation.

For a polytropic model of the solar convection zone, the unperturbed eigenmodes are sums of products of exponentials and powers (PEP). We have chosen, Eq.(38), a convection stream function which is also a PEP. The perturbation theory requires a number of differentiations, multiplications and additions (which leave a PEP a PEP), followed by an integration (a trivial algebraic operation for a PEP). We wrote a Numerical Algebra program which performs all these operations. The program was tested as follows. In the interval $0.1<k/q<0.6$ the calculated frequency shifts coincided with the shifts calculated by the direct numerical simulations program described in Appendix B. Another reason to believe that we did not make a mistake, is that we succeeded in reconstructing analytical formulae with a correct algebraic structure from our Numerical Algebra program (Appendix C).

We now present the results. For a single Fourier harmonic of the flow, the f-mode fractional frequency shift is shown in Fig.1. 
The long-wavelength ($k\rightarrow 0$) shift is positive. The poles in the vicinity of $k/q=0.7$ are due to resonant scattering. The resonant condition, $\omega _{n}^2(k\pm q)=\omega _0^2(k)$, is satisfied by two families of solutions. The first family is
\begin{equation} 
k_n={2n+m\over 2n+2m}q,
\end{equation}
which in our case ($m=4$) gives resonances at $k/q=(n+2)/(n+4)=$ 0.5, 0.6, 0.67, 0.71, 0.75, 0.78, 0.8, ... The first two of these are actually absent because the corresponding matrix elements vanish. The next five are seen in Fig.1, and higher resonances are not present because we summed the second-order PT shift, Eq.(35), only up to $n'=6$. 

The second family of resonances is given by
\begin{equation}
k_n=(1+{m\over 2n})q,
\end{equation}
which in our case is $k/q=1+2/n=$ 3, 2, 1.5, ... These are shown in Fig.2, where the ``red'' resonances are too weak to be seen. Fig.2 also shows that the resonances disappear without a trace when integrated over a spectrum of turbulent modes. Even a rather narrow spectrum used in Fig.2 reduces a complicated pattern of resonances to just two asymptotics and a transition region between them.

The short-wavelength (high-degree) frequency shift does not depend on the aspect ratio of the turbulent eddies, $\mu$, nor on the turbulence wavenumber, $q$. With an accuracy better than 1\% , the high-degree shift is given by
\begin{equation}
s_{uv}=-M^2,
\end{equation}
where $M$ is the surface Mach number of the turbulence. The low-degree shift depends on $\mu$ and $q$. It is given by the following exact formula
\begin{equation}
s_{ir}={5\over 242\mu ^6}M^2(k/q)^4,
\end{equation}
which is applicable for $k/q\ll \mu$. Smaller aspect ratios $\mu$ give higher maximal positive shifts $s_{max}$, but we find that even the very small $\mu=0.1$, gives  $s_{max}\approx 0.1\% $, for a typical value of the Mach number, $M=0.1$. This is smaller than the current observational accuracy.

The smallness of the maximal positive shift may be explained as follows. At very small $k$, the mode's energy is concentrated at large depths, where the turbulence is too weak to produce a noticeable frequency shift. At larger $k$, the mode-turbulence interaction becomes stronger, but the negative short-wavelength shift takes over.

The f-mode frequency shift is practically independent of the polytropic index $m$. The high-degree shift for $m=2$ is $s_{uv}=-1.1M^2$, which is close to the $m=4$ result $s_{uv}=-M^2$. We also calculated the frequency shifts for the $p_1$-mode and obtained similar results.

The shape of the spectrum-averaged curve in Fig.2. seems to be correct. Shown in Fig.3 are the data of Libbrecht, Woodard \& Kaufmann (1990) and the theoretical curve of Fig.2 scaled to match the observations. In physical terms, the required scaling corresponds to the Mach number of the turbulent convection $M\sim 0.3$, and the turbulent length scale $\lambda \sim 3000km$.

\section{CONCLUSION}

We have calculated the f-mode frequency shifts due to scattering of waves by turbulence using perturbation theory. A two dimensional model of convection was employed, but our methods are applicable to a three dimensional problem with only minor changes. Given a turbulent velocity field, one can use the perturbation theory to calculate the turbulent f-mode frequency shifts and line-widths. The main qualitative results for our simplified model are that the long-wavelength (wavelength larger than the turbulent eddy size) frequency shifts are positive but unobservably small and the short-wavelength fractional shifts are negative and of the order of $M^2$. 

\acknowledgements

During the preparation of this manuscript, I received help and encouragement from Pawan Kumar, who suggested the problem. 

I thank John Bahcall, Daniel Eisenstein, Peter Goldreich, Michael Isichenko, and Eli Waxman for valuable discussions. This work was supported by NSF PHY-9513835.

\appendix

\section{A Simple Shear Flow}
To test the homogeneous flow results, Eqs. (11), (14), consider sound wave propagation in a simple two-dimensional shear flow. Let the flow velocity along the $y$-axis be $+V$ in the intervals $2n<x<2n+1$, and $-V$ in the intervals $2n-1<x<2n$, where $n$ is an integer. Consider a sound wave with a wavenumber $k$ propagating in the $y$ direction. An exact dispersion relation can be  obtained by usual methods. It reads 
\begin{equation}
{k_1\tanh k_1a\over (\omega -kV)^2}={k_2\tan k_2a\over (\omega +kV)^2},
\end{equation}
where $a=1/2$ is the half-width of the layers, and $k_{1,2}$ are defined by
\begin{equation}
(\omega \pm kV)^2=c^2(k^2\pm k_{2,1}^2).
\end{equation}
These equations can be solved analytically in the second order in $V$, the result is
\begin{equation}
\delta \omega ^2=V^2(3k^2-{1\over3}k^4).
\end{equation}
On the other hand, Eqs. (11), (14) predict
\begin{equation}
\delta \omega ^2=V^2(3k^2-4<{1\over q^2}>k^4).
\end{equation}
Since the average inverse squared wavenumber of the flow is
\begin{equation}
<{1\over q^2}>={1\over \pi ^2}\sum (2n+1)^{-4}/\sum (2n+1)^{-2}={1\over12},
\end{equation}
the two results agree. 

\section{Numerical Simulations}

In two dimensions, for a single horizontal Fourier harmonic of the turbulent flow $V$, direct numerical solution of the eigensystem describing linear sound waves is possible. If we wish to determine the quadratic in $M$ frequency shift of an acoustic wave with a horizontal wavenumber $k$, only two satellites of the wave, $k-q$ and $k+q$ ($q$ is the turbulence wavenumber) must be included into the calculation. A system of ordinary differential equations results as described below. 

Write the system (25), (26) as
\begin{equation}
P_0\Phi _k=P_q\Phi _{k-q}+P_{-q}\Phi _{k+q}.
\end{equation}
Here $\Phi _{k}=(v_x,v_y,\rho )$ is the acoustic wave, the subscript $k$ refers to the horizontal wavenumber. The unperturbed operator $P_0$ does not depend on the flow $V$, Eq. (B1) with a zero right hand side (RHS) gives an eigenmode in the absence of the flow. The perturbation operator $P_q$ is linear in $V$, the subscript $q$ denotes the horizontal wavenumber of the flow $V$. To within the $M^2$ accuracy, the satellites satisfy 
\begin{equation}
P_0\Phi _{k+q}=P_q\Phi _{k},
\end{equation}
\begin{equation}
P_0\Phi _{k-q}=P_{-q}\Phi _{k}.
\end{equation}

Equations (B1)-(B3) were solved in three steps: (i) the RHS in (B1) is zero, shoot to determine the unperturbed mode $\Phi _k$. (ii) use the unperturbed mode $\Phi _k$ in the RHS of (B2), (B3), shoot to determine the satellites $\Phi _{k\pm q}$. (iii) use both satellites in the RHS of (B1), shoot to determine the corrected mode $\Phi _k$. Solution obtained in (iii) should agree with the boundary conditions at the surface, $x=0$, and at the bottom, $x=H$.

In the absence of stratification ($g=0$, $c=const$, $\rho _0=const$), for a number of various flows $V$, our numerical results agreed with the theory of \S 2. For the solar stratification, reliable results were obtained only for $0.1<k/q<0.6$. The reasons for a poor performance of the program outside of this interval seem to be as follows. Small $k$ require too large depths $H$, because the mode is cut off at about $4/k$. Larger wavenumbers, $k\sim q$, are in the resonance region, and the resonances are sensitive to the position of our unphysical bottom $H$. However, numerical frequency shifts in the interval $0.1<k/q<0.6$ were $H$-independent to a few percent accuracy. These results agreed with the exact (up to the numerical algebra) calculation used in \S 3.

\section{Exact Results}

Knowing the algebraic structure of the frequency shifts formulae, we used the Numerical Algebra program to obtain exact results. The simplest example is the first order PT shift $\delta _2\omega _0^2(k)$ given by Eq.(34). This shift is an integral of a sum of products of exponentials and powers (PEP). Rescale the length, so that the turbulence wavenumber is $q=1$, also take a unit-amplitude convection, A=1. Then the exponential function of the PEP under the integral (34) is $e^{-2(k+\mu )x}$. The highest power of $x$ in the PEP is 7. Therefore, the resulting shift must be a polynomial in $k$ and $\mu$  divided by $(k+\mu )^8$. The Numerical Algebra confirms this result. The reconstructed exact first order PT shift (34) is given by 
\begin{equation}
s_2={\delta _2\omega \over \omega }={1\over 4}{k^4\over (\mu +k)^8}(a_0+a_1k+a_2k^2+a_3k^3+a_4k^4),
\end{equation}
where
\begin{equation}
a_0=20\mu ^2,
\end{equation}
\begin{equation}
a_1=-(20\mu +82\mu ^3),
\end{equation}
\begin{equation}
a_2=-(250+191\mu ^2),
\end{equation}
\begin{equation}
a_3=44\mu ,
\end{equation}
\begin{equation}
a_4=363.
\end{equation}
We used our Numerical Algebra Program to reconstruct analogous formulae for the second order PT sum, Eq.(35). These exact formulae were used to integrate out the resonances, as shown in Fig. 2. The matrix elements $<n,k-q|Q_1|0,k>$ are identically zero for $k<q$ and $n=0,1$. This explains the absence of ``red'' resonances at $k/q=0.5$ and at $k/q=0.6$ as seen in Fig.1.

The lowest order in $k$ long-wavelength shift is entirely due to the first order PT shift, Eq.(C1). As it is, Eq. (C1) gives an asymptotic fractional shift $s_{ir}=5k^4/\mu ^6$. The stream function amplitude $A=1$ corresponds to the surface Mach number $M=\sqrt{m}(m+3/2)=11$, for both of the Fourier harmonics $q=\pm 1$. This gives the mean squared Mach number, $M^2=242$, explaining the origin of 242 in the denominator of Eq. (42).

\newpage

\figcaption{Fractional frequency shift ($s$) normalized by the turbulence Mach number squared ($M^2$) as a function of the  wavenumber ($k$) normalized by the turbulence wavenumber ($q$).}

\figcaption{Resonances of the frequency shift are removed by averaging over a spectrum of turbulence wavenumbers. The spectrum was $A_q^2\sim q^4 \exp (-2q^2)$ with a fixed aspect ratio $\mu =1$.}

\figcaption{Fractional frequency shift. The theoretical curve of Fig.2 was scaled to match the data of Libbrecht et al (1990) as explained in the text.}

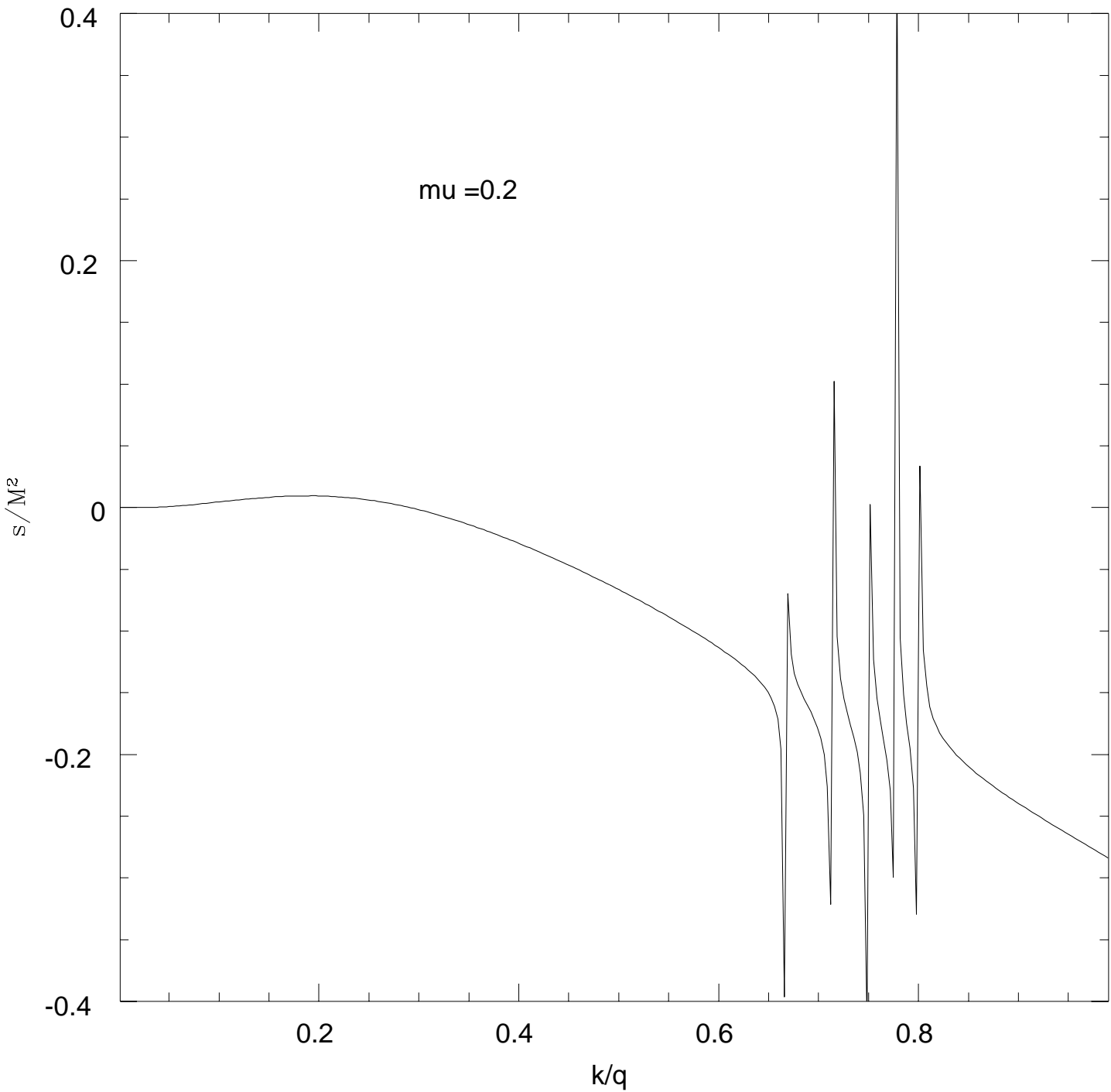
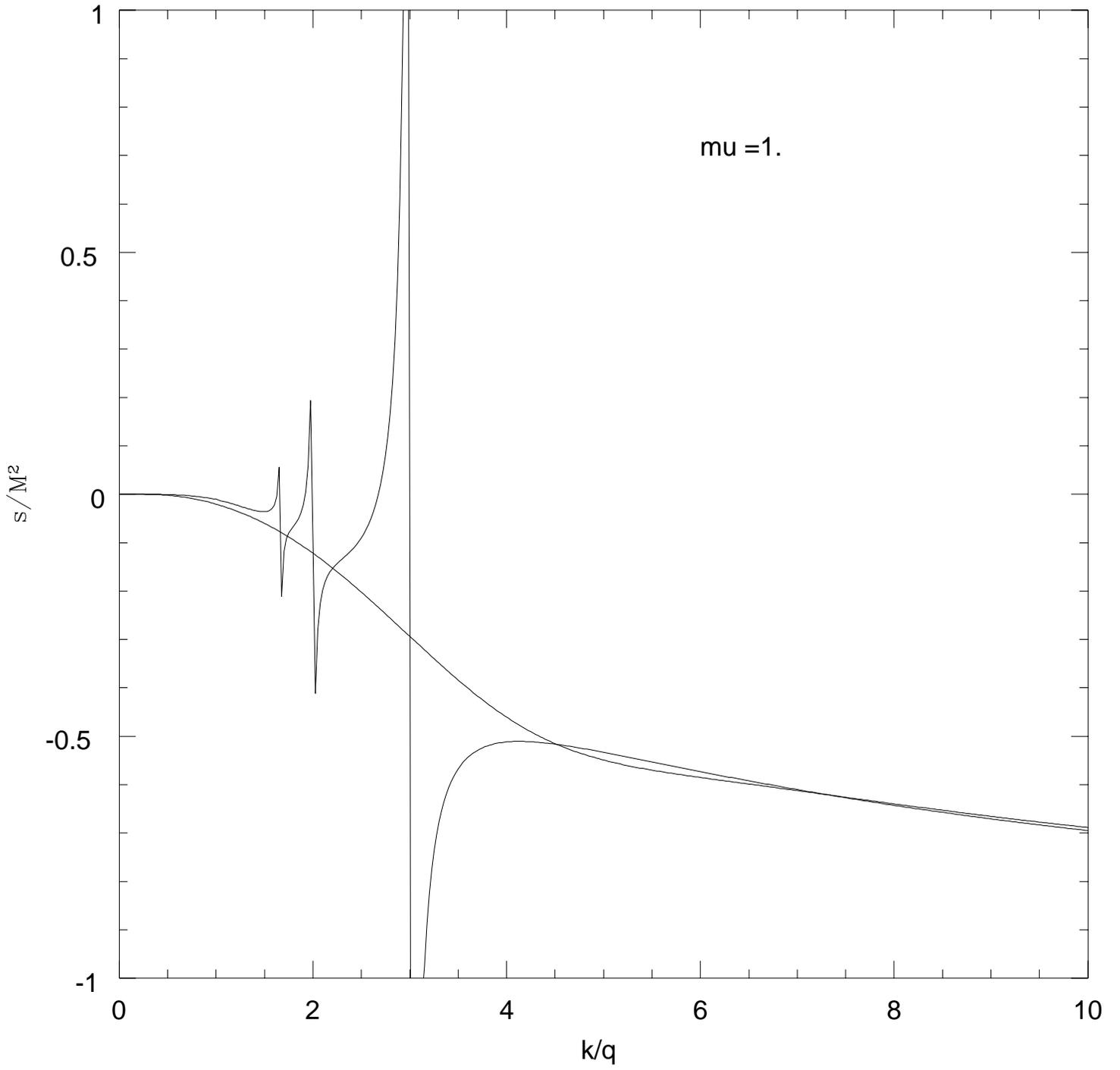
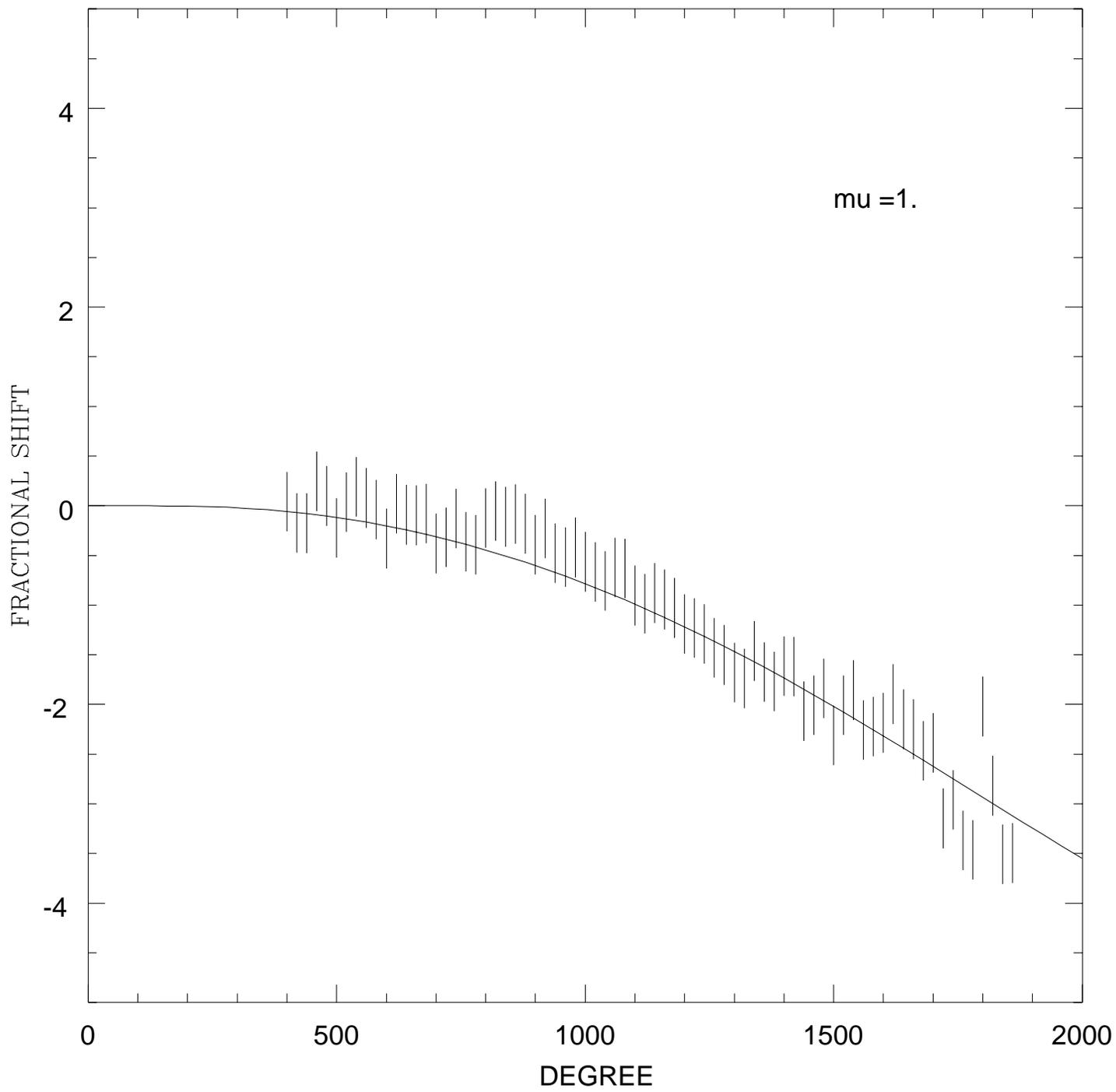
\end{document}